\documentclass[12pt]{article}

\usepackage{latexsym,amsmath,amssymb,theorem,epsfig}

\usepackage{graphicx}
\usepackage{graphicx,subfigure}
\usepackage{epstopdf}
\usepackage[body={17.5cm, 21cm},right=2cm]{geometry}
\usepackage{amssymb}
\usepackage{amsmath}
\usepackage{psfrag}
\usepackage{epsfig}
\usepackage{cancel}
 \allowdisplaybreaks[4]

\usepackage[all]{xy}

\DeclareFontFamily{OT1}{rsfs10}{}
\DeclareFontShape{OT1}{rsfs10}{m}{n}{ <-> rsfs10 }{}
\DeclareMathAlphabet{\mathscript}{OT1}{rsfs10}{m}{n}

\numberwithin{equation}{section}

\newcommand{\ns}{\normalsize}

\newcommand{\be}{\beta}

\def\gsim{ \lower .75ex \hbox{$\sim$} \llap{\raise .27ex \hbox{$>$}} }
\def\lsim{ \lower .75ex \hbox{$\sim$} \llap{\raise .27ex \hbox{$<$}} }
\def\be{\begin{equation}}
\def\ee{\end{equation}}
\def\bea{\begin{eqnarray}}
\def\eea{\end{eqnarray}}

\begin{document}

\begin{titlepage}


\title{
   {\LARGE   Infrared-modified Universe\footnote{Essay written for the Gravity Research Foundation 2012 Awards for Essays on Gravitation.}}
}

\author{
   Federico Piazza\footnote{fpiazza@apc.univ-paris7.fr}
     \\[0.5em]
   {\ns PCCP and APC, Universit\'e Paris 7, 75205 Paris, France}\\[1cm]}

\date{}

\maketitle

\begin{abstract}
We consider a Hubble expansion law modified in the infra-red by distance-dependent terms, and attempt to enforce homogeneity upon it. As a warm-up, we re-derive the basic kinematics of a Friedman Robertson Walker universe without using standard general relativistic tools:  we describe the expansion with a `Hubble velocity field' rather than with a four dimensional metric. Then we extend this analysis to the modified Hubble expansion and impose a transformation for velocities that makes it identical for all comoving observers, and therefore homogeneous.  We derive the modified equation for light ray trajectories and other geometrical properties that are incompatible with the general relativistic description. We speculate that this extended framework could help addressing cosmological problems which are normally explained with accelerating expansions.

\end{abstract}

\thispagestyle{empty}

\end{titlepage}

\section{Introduction}

Cosmological data provide strong evidence for two mechanisms beyond the minimal big-bang model: `inflation' and `dark energy'. 
In a broad sense, `inflation' is whatever makes the universe homogeneous and correlated on large scales; `dark energy' what corrects the luminosity distance for its observed redshift-dependence~\cite{ruth}.
One might well suspect these two phenomena to be related. After all, taken at face value, both the high redshift behavior of luminosity distances and the temperature fluctuations entering our causal horizon concern the infrared physics of Hubble scales and beyond: why do not try to give them a unified explanation? 
Such an `Okham's razor' attitude clashes with general relativity (GR). In a homogeneous universe, luminosity distance and expansion are related in a very specific way. It follows that dark energy must be very recent, or the successful expansion history since, say, nucleosynthesis would be spoiled. On the other hand, the decelerating phases of radiation and matter domination cannot explain homogeneity: we need to assume \emph{another} epoch of acceleration to happen before the onset of deceleration, at much earlier times and higher energies.

Still, it looks curious having to invoke accelerating expansion \emph{twice}. Before definitely embracing the more and more successful `DE-CDM+Inflation' paradigm, it is worth asking whether some theoretical framework beyond GR could suggest a unified explanation, a dramatic simplification. The task looks challenging because causality and expansion are entangled in GR at the pure kinematical level, \emph{i.e.}, just by the form of the Friedman Robertson Walker (FRW) line element. Models of massive gravity, like any other modification of the Einstein equations (see e.g.~\cite{costas}), do not impact the basic kinematics of a FRW universe. The infra-red modification that we are after needs to be \emph{geometrical} in the first place, because it has to disentangle causality and luminosity distances from the expansion history. 

In this paper we explore a geometrical modification of the FRW paradigm at large scales defined by a modified expansion law for the comoving observers which also depends on their distances,
\begin{equation} \label{basic}
\dot X = H X \left[1+g(X;t)\right] .
\end{equation}
In the above, $X(t)$ is the distance between any pair of comoving observers, a dot means derivative with respect to proper time, $H(t)$ is the Hubble parameter and $g$ a function that starts quadratically in $X$. Note that in any FRW universe the separation among comoving observers grows proportionally to the scale factor $a(t)$, and therefore $g = 0$.  In this paper we try to reconcile~\eqref{basic} with homogeneity and make sense of it being valid for \emph{any} pair of comoving observers. 

This approach is suggested by the `ultra-strong', or `extreme' equivalence principle (EEP)~\cite{usep1}. EEP postulates that there is a vacuum state for test fields on which the energy momentum expectation value is the same as in flat space (i.e. local and non-local space-time dependent contributions to $\langle T_{\mu \nu}\rangle_{\rm bare}$ are absent), and that the geometry of spacetime as described in GR should be modified accordingly.
Stationary spacetimes (e.g. Minkowski and AdS) already allow such a state and therefore are immune from the invoked modification. On the opposite, EEP suggests that time dependent spacetimes as described in GR are only local approximations of the `real theory'. When applied to a minimally coupled scalar field in a FRW universe, EEP can be enforced~\cite{usep2} by a modified dispersion relation, that in real space translates into a modified expansion law of the above type with 
\begin{equation} \label{basic2}
g(X;t) = \frac{X^2}{2}(\dot H + H^2) + {\cal O}(X^4).
\end{equation}
Such a modification is infra-red because it starts being effective at Hubble scales. Here we show that its impact on cosmology can be stronger than appeared from earlier analyses~\cite{usep2,usep3}.

\section{FRW without GR}

In GR spatial homogeneity and isotropy are realized by imposing a group of isometries on the metric tensor, that univocally lead to the FRW metric.
In this section we reconstruct some properties of a FRW Universe by using looser ingredients than the metric. This will allow us in the next section to generalize the concept of cosmological homogeneity beyond GR. 

In a spatially flat FRW, at every value $t$ of her proper time, a co-moving observer describes the set of all simultaneous events as a 3-dimensional Euclidean space $\mathbb R^3$.  At a later time, the expansion of the universe has taken all comoving observers further apart. 
The observer at the origin represents this state of affair with the \emph{Hubble velocity field}, $\vec h(\vec X)$, representing the Hubble flow at position $\vec X$ (both $\vec X$ and $\vec h$ are 3-dimensional vectors). The position of the comoving observer $\vec X$ after a time interval $d t$ is 
\begin{equation} \label{shift}
\vec X(t+ d t) = \vec X(t) + \vec h(\vec X(t))\  d t\, .
\end{equation}
In a FRW universe the Hubble velocity field $\vec h(\vec X)$ is proportional to the distance,
\begin{equation} \label{frw_hubble}
\vec h(\vec X) = H \vec X\, ,
\end{equation}
where $H=\dot a/a$ is the Hubble parameter. 

The mapping~\eqref{shift}-\eqref{frw_hubble} appears centered around the origin. However, homogeneity can be enforced by assigning a transformation law for the velocity fields when they are ``seen" by different observers. 
In GR such a transformation is Galilean: velocities are simply added to the Hubble flow. A velocity field $\vec v (\vec X)$ `seen by' the point $\vec A$ reads
\begin{equation} \label{frw_trans}
\vec v_{\vec A} (\vec X) =  \vec v (\vec X) - H \vec A\, ,
\end{equation}
where the index means ``as seen from" -- no index implies ``as seen from the origin". 
This transformation can be especially applied to the Hubble-velocity field $\vec h(\vec X)$, giving
\begin{equation} \label{homo}
\vec h_{\vec A} (\vec X) = \vec h(\vec X - \vec A)\, . 
\end{equation}
The above crucial relation is what defines homogeneity in this framework: the expansion looks the same for any observer. 
As a corollary, $h_{\vec A}(\vec A) = 0$ as expected. 

Note that, in this description, what happens at another point might look totally unphysical. For example, 
the Hubble velocity at $X > H^{-1}$ (beyond the cosmological horizon) is superluminal. What needs to be at most unity, because directly measurable, is the velocity at a point $\vec X$ \emph{seen} at the same point $\vec X$: $v_{\vec X}({\vec X}) \leq 1$. 
This is the element of \emph{general} (rather than special-) relativistic physics in this construction. 
We will call $v_{\vec X}({\vec X})$, \emph{local at} $\vec X$.
By definition, a light ray has always unit local velocity and thus is defined as a function $\vec L(t)$ satisfying $d \vec L_{\vec L}(t)/dt = 1$. By using~\eqref{frw_trans} we obtain the correct equation for a light ray,
\begin{equation} \label{GRlight0}
\frac{d L}{d t} = 1 + H L\, .
\end{equation}
By switching to conformal coordinates $\vec l = \vec L/a(t)$ and time $d \tau = dt/a(t)$, we recover the familiar
\begin{equation} \label{GRlight}
\frac{d l}{d \tau} = 1 .
\end{equation}
Given an expansion history $a(t)$, the trajectories of comoving observers~\eqref{frw_hubble} and light rays~\eqref{GRlight0} define the basic kinematic and causality of a FRW universe. The descriptions that different comoving observers give are related by the transformation law for velocities~\eqref{frw_trans} and by simple space translations. Nowhere in above construction is the concept of four-dimensional metric used.

\section{Beyond FRW}

We now turn to a more general Hubble velocity field than~\eqref{frw_hubble}, such as that of equation~\eqref{basic},
\begin{equation} \label{hubble}
\vec h(\vec X) = H \vec X \left[1+ g(X;t)\right] \, .
\end{equation}
We recall that $g$ is a function that starts quadratically in $X$ and that, according to the EEP~\cite{usep1,usep2}, it is given by~\eqref{basic2}.
While keeping $g$ general whenever possible, we find the absence of mass parameters in~\eqref{basic2} very appealing; therefore, we assume at least that $X$ will appear in $g$ always multiplied by some power of $H$ and its derivatives, rather than by some given mass parameter $m$. If we assume a constant equation of state, this means that $g$ is in fact a function of $(X/t)^2$. 

Since we insist that the universe be homogeneous, we want to enforce a transformation law for velocities replacing~\eqref{frw_trans} such that the modified Hubble expansion looks the same at any point, which is our definition of homogeneity. Unfortunately, such a transformation is not unique. Here we consider what looks like the simplest and most reasonable choice,
\begin{equation} \label{trans}
\vec v_{\vec A} (\vec X) = \left(\frac{\vec v (\vec X)}{1 + g(X)}  - H \vec A\right)\left(1+ g(|\vec X - \vec A|)\right)\, ,
\end{equation}
but we should warn the reader that, to some extent, the implications that we derive later depend on such a choice. In a more symmetric fashion,  
\begin{equation}
\frac{\vec v_{\vec A} (\vec X)}{1+ g(|\vec X - \vec A|)} - \frac{\vec v_{\vec B} (\vec X)}{1+ g(|\vec X - \vec B|)} = H(\vec B - \vec A)\, .
\end{equation}
From~\eqref{trans}, the transformation law of a \emph{local} velocity vector at $\vec A$ as seen from the origin is found:
\begin{equation}\label{translocal}
 \vec v (\vec A) = \left(1 + g(A)\right)  \left(H \vec A  + \vec v_{\vec A} (\vec A) \right) \, .
 \end{equation}
 
 Similarly to the GR case, we define null rays as those having always unit local velocity. So they satisfy the equation
 \begin{equation}
 \frac{d L}{dt} = (1+HL)\left(1+g(L)\right)\, .
 \end{equation}
 
 We can, again, switch to `comoving coordinates' $\vec x = \vec X/a(t)$, that we denote here with lower-case latin letters. However, notice that in this modified framework comoving observers are no longer charactrized by $\vec x = const.$. Rather, according to~\eqref{hubble},
\begin{equation} \label{anomalous}
\dot {\vec x} = \vec x H g(a x)\, .
\end{equation} 
By using comoving coordinates and conformal time the equation of a light ray can be written as 
 \begin{equation} \label{light}
\frac{d\, l(\tau)}{d \tau} = 1 + g(L)(1 + H L ),
\end{equation}
where $L = a(\tau) l(\tau)$. Without taking into proper account the implications of homogeneity, in~\cite{usep2,usep3} the equation of a light ray was written by simply adding the anomalous piece~\eqref{anomalous} to the usual GR one---eqs.~\eqref{GRlight0} and~\eqref{GRlight}.
 
\section{Geometrical Considerations}
\begin{figure}[htbp] 
\begin{center} \vspace{-.5cm}
\includegraphics*[width=3.7in]{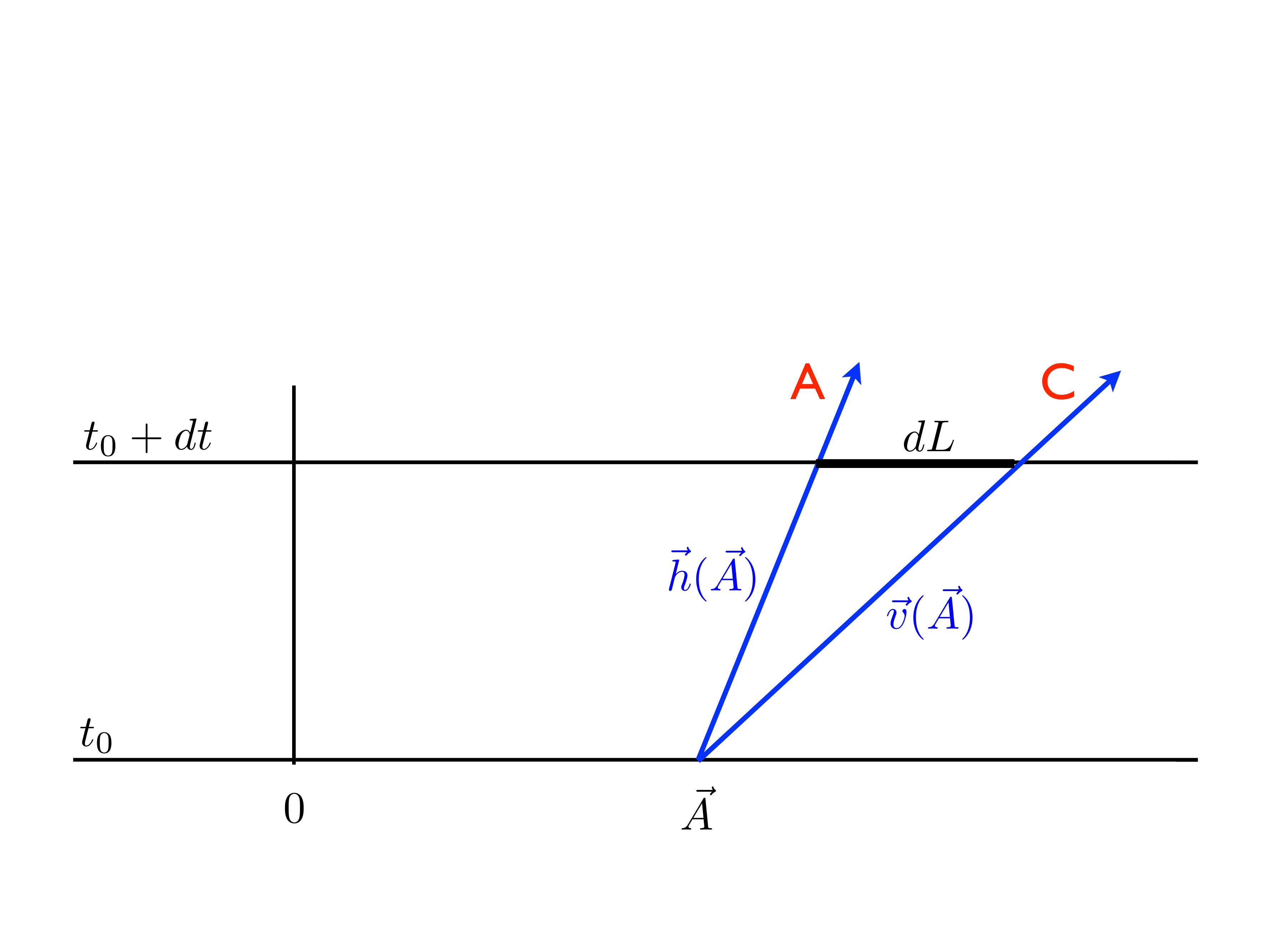}\!\!\!\!\!\!\!\!\!\!\!\!\!\!\!\!\includegraphics*[width=3.7in]{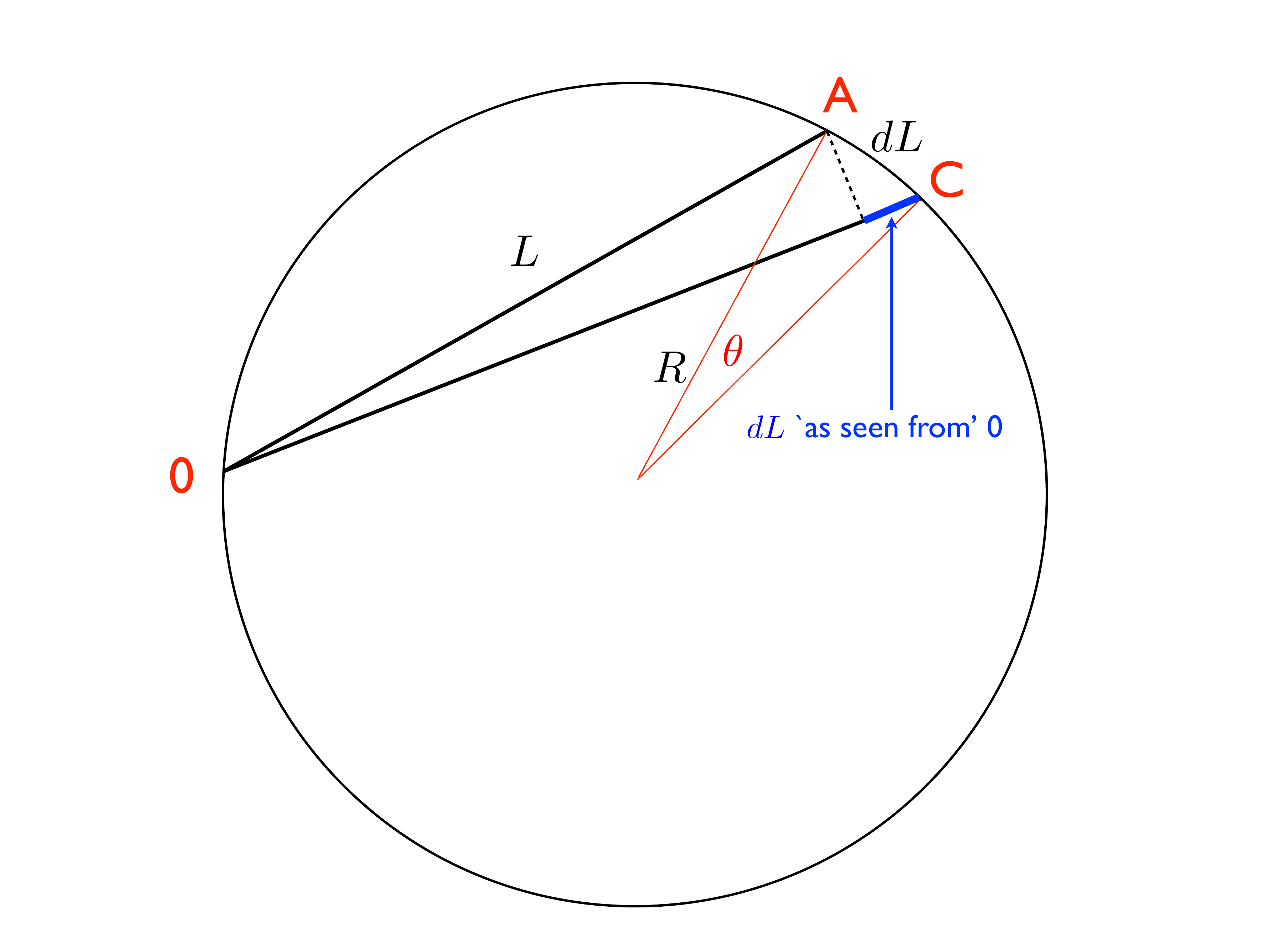}%
\end{center}\vspace{-.5cm}
\caption{\small Left panel: the situation of two observers $A$ and $C$ described in the text: $A$ is comoving -- its velocity vector is $\vec h(\vec A)$, while $C$ has velocity $\vec v(\vec A)$. The non-linearity of the velocity transformation~\eqref{translocal} makes the length interval $d L$ dependent of the observer. Right panel: a simple pictorial way to make sense of a distance that is not additive and of a metric space that is  a Riemannian manifold only ``\emph{locally}".}
\label{figure1} \end{figure}

The non-linearity of velocity transformations~\eqref{trans} has inevitable geometrical consequences. Consider two observers $A$ and $C$ (Figure~\ref{figure1}, left panel) described by trajectories $\vec A(t)$ and $\vec C(t)$. $A$ is comoving while $C$  is not. Say that, at some time $t_0$, $C$ intersects $A$, $\vec C(t_0) = \vec A(t_0)$, with some local velocity
$\vec v_{\vec A}(\vec A)$. This means that at a later time $t_0+dt$, the distance between $A$ and $C$ is $ d\vec L = \vec v_{\vec A}(\vec A) dt$. Let us see how the same situation is seen from the faraway origin. The velocity of $C$ can be read from~\eqref{translocal}. By subtracting the velocity of $A$, the distance between $A$ and $C$ at $t_0+dt$  is calculated as $ \vec v_{\vec A}(\vec A) (1+ g(A))dt$.  We get to the puzzling conclusion that the same line element $d\vec L$, when accounted for from a distance $X$, gets rescaled by a conformal factor $1+g(X)$:
\begin{equation}\label{puzzle}
d\vec L(`{\rm seen\ from\ distance}\ X ') = (1+ g(X))d\vec L\, .
\end{equation}

In one dimension we can intuitively make sense of this puzzling behavior in a simple way. Consider a circle ${\cal C}$ of radius $R$ (Figure~\ref{figure1}, right panel) and define the distance between any two points as the length of the \emph{chord} (rather than the arc!) between them: two points at a relative angle $\theta$ have distance $L(\theta) = 2 R \sin(\theta/2)$. Equipped with such a distance, ${\cal C}$ is a metric space; it is `homogeneous', in the sense that there is a set of transformations (rotations) that preserve the mutual distances between the points. However, ${\cal C}$ is not a Riemannian manifold, as there is no line element that, integrated, gives the chosen distance $L$. It is tempting to say that this metric space is `\emph{locally}' a metric manifold, because, at small angles, $L$ and the Riemannian distance $\theta R$ coincide.

Going back to the puzzle of our two observers, we are lead to conclude that on our modified FRW universe distances are not additive and therefore cannot be expressed as an integral of a line element. By looking at the example of the circle, there is a simple way we can make sense of~\eqref{puzzle} in one dimension. Since $A$ and $C$ are very close to each other, $dL = R d \theta$. On the other hand, if we attempt to define the distance between $A$ and $C$ as a \emph{difference} of distances, we differentiate $L(\theta)$ and get something analogous to~\eqref{puzzle}, 
\begin{equation}
dL({\rm seen\ from\ distance}\ L) = R \cos\left(\frac{\theta}{2}\right) d\theta =  \left(1 - \frac{L^2}{8 R^2} + \dots\right) dL\, .
\end{equation}

In 3 dimensions it is not clear how to make sense of~\eqref{puzzle} in such a pictorial way. 
In particular, it is not clear whether or not, beside the abstract notion of distance between two points, we can also consistently associate a length with any space-like curve, nor whether that is needed. 
Giving up the `principle' of additivity of space-like distances may look grotesque.
On the other hand, any firm operational basis for such a principle, as far as we have been able to see, seems rooted in the possibility of repeating some measurement operation (e.g. laying down a ruler) many times in the same physical conditions. This looks possible only in stationary spacetimes, that are in fact immune from the EEP's modification. While leaving to future work a better understanding of various issues related to this modified geometry, here we content ourselves with the mnemonic rule~\eqref{puzzle} which, together with equation~\eqref{light}, is the main message of this paper. 

\section{Outlook}

In this note we considered an infra-red modification of the FRW model that corrects the usual expansion law $\dot X = H X$ by distance-dependent terms (eq.~\ref{basic}), and tried to make it consistent with homogeneity. We derived a modified equation for the light rays~\eqref{light}. Perhaps more importantly, we found that the clash between the modified expansion~\eqref{basic} and homogeneity can be reconciled at the expense that space-like distances  are no longer additive on time-dependent spacetimes. In order to illustrate this point  we used the one-dimensional example of a circle, with distances defined by chords rather than arcs. Non additivity of distances looks incompatible with a Riemannian manifold, which is at the basis of the GR description of physical events. 

As radical as it is, this approach contains potential pay-offs and opportunities. First, the proposed modification impacts the calculation of any cosmological distance without changing the (local) expansion history. The mnemonic rule~\eqref{puzzle} suggests, roughly, that distances calculated in GR should in fact be divided by $1+g(z)$. Since $g$ defined in~\eqref{basic2} is negative in a decelerating universe, this prescription amplifies cosmological  distances and goes in the direction of an effective acceleration. Second, the causality and horizon structure of a FRW universe is also modified, again, without the local expansion being touched. It will be interesting to see how comoving volumes are `swept' by the light rays according to the modified equations that we have derived. Whether or not these effects can address dark energy and inflation, they are permanently at work and active at space-like separations of order Hubble, rather than being `happening' at given times and energy scales. 
Finally, the proposed modification~\eqref{basic}, \eqref{basic2} does not contain any adjustable mass parameter, and the numerical ones can be fixed~\cite{usep2} by enforcing the EEP. While the prevalent model-building attitude demands at least one new parameter for each cosmological problem to solve -- but potentially many more -- this framework does not contain in principle any more free parameter than GR itself.  Had the universe done us the favor of being simple, and not merely beautiful, venturing into such an uncommon theoretical set-up could prove rewarding. 

\section*{Acknowledgements}

I thank George Smoot for his comments on the manuscript, including recommending a new name -- `extreme' rather than `ultra-strong' -- for the principle originally proposed in~\cite{usep1}. I also thank Ruth Durrer, Francesco Nitti, Daniele Steer and Filippo Vernizzi.

\end{document}